\documentclass[12pt]{ronbun}

\usepackage{color,amsmath,amssymb,graphicx}

\usepackage{cite}
\usepackage{bm}
\usepackage{dcolumn}

\newcommand{\s}{\scriptscriptstyle}

\newcommand{\pd}{\partial}

\newcommand{\nn}{\nonumber}
\newcommand{\e}{{\rm e}}

\newcommand{\del}{\delta}

\newcommand{\rar}{\rightarrow}

\newcommand{\al}{\alpha}

\newcommand{\ba}{\begin{eqnarray*}}
\newcommand{\ea}{\end{eqnarray*}}

\newcommand{\up}{\uparrow}
\newcommand{\down}{\downarrow}

\newcommand{\KUCPlogo}{\hbox{\lower 1.4ex\hbox{
\Huge\boldmath $\cal K$}
\kern -1.15em {\sffamily \bfseries\large\ UCP}}
\kern -4.5em \raise 0.2em\hbox{\lower 1.4ex\hbox{\color{cyan}
\Huge\boldmath $\cal K$}
\kern -1.15em {\color{magenta}\sffamily \bfseries\large\ UCP}
\put(-20,-7){\tiny\it preprint}
}}

\numberwithin{equation}{section}

\begin{document}

\begin{flushright}

\parbox{3.2cm}{
{KUCP-0217 \hfill \\
{\tt hep-th/0209081}}\\
\date
 }
\end{flushright}

\vspace*{0.7cm}

\begin{center}
 \Large\bf Ground State  of Supermembrane on PP-wave
\end{center}

\vspace*{1.0cm}

\centerline{\large Noriko Nakayama$^{\dagger a}$, 
Katsuyuki Sugiyama$^{\ast}$ and Kentaroh Yoshida$^{\dagger b}$}

\begin{center}
$^{\dagger}$\emph{Graduate School of Human and Environmental Studies,
\\ Kyoto University, Kyoto 606-8501, Japan.} \\
{\tt E-mail:~$^{a}\,$nakayama@phys.h.kyoto-u.ac.jp \\
\qquad\quad\,  $^b\,$yoshida@phys.h.kyoto-u.ac.jp} \\
\vspace{0.2cm}
$^{\ast}$\emph{Department of Fundamental Sciences, \\
Faculty of Integrated Human Studies, \\
Kyoto University, Kyoto, 606-8501, Japan.} \\
{\tt E-mail:~sugiyama@phys.h.kyoto-u.ac.jp}
\end{center}

\vspace*{1.0cm}

\centerline{\bf Abstract}

We consider the ground state of supermembrane on the maximally
supersymmetric pp-wave background 
by using the quantum-mechanical procedure of de Wit-Hoppe-Nicolai.  
In the pp-wave case the ground state has non-trivial structure 
even in the zero-mode Hamiltonian, which 
is identical with that of superparticles on
the pp-wave 
and resembles supersymmetric harmonic oscillators. 
The supergravity multiplet in the flat case comes to split with a certain
energy difference.  We explicitly construct 
the unique supersymmetric ground-state wave function of the zero-mode 
Hamiltonian, which is obviously normalizable. 
Moreover, we discuss the nonzero-mode Hamiltonian 
and construct an example for the 
ground-state wave function with a truncation of the variables. 
This special solution seems non-normalizable but its $L^2$-norm can be 
represented by an asymptotic series.

\vspace*{1.2cm}

\thispagestyle{empty}
\setcounter{page}{0}

\newpage 

\tableofcontents 

\section{Introduction}

Recently, strings and M-theory on the pp-wave have been 
greatly focused on. 
The maximally supersymmetric 
pp-wave background in eleven dimensions \cite{KG} (KG solution) is one of the 
candidates for M-theory background. This background is achieved 
from the geometry of 
$AdS_{4}\times S^7$ or $AdS_{7}\times S^4$ \cite{OP} via the 
Penrose limit \cite{Penrose}. Similarly, the maximally supersymmetric 
type IIB pp-wave 
background was found \cite{OP2}, on which the 
string theory is exactly solvable \cite{M,MT,RT}. 
Moreover, this string theory was used in the context of 
the $AdS$/CFT correspondence between the string spectra and 
the operators in Yang-Mills side \cite{Malda}.  

Furthermore, the matrix model on the pp-wave was proposed 
in Ref.\,\cite{Malda}. 
Originally, its action is obtained from the study of 
superparticles. In the flat space, the supermembrane theory 
\cite{sezgin,bergshoeff,dWHN} leads to the matrix model through the 
matrix regularization \cite{dWHN}, 
which is considered as a candidate for M-theory \cite{BFSS}. Following the same
procedure, the matrix model on the pp-wave can be also 
obtained from the supermembrane \cite{DSR}. 
It was also shown that the superalgebra of the matrix model agrees 
with that of the supermembrane \cite{SY} as in 
the flat case \cite{BSS}. 

In our previous works, we studied the supermembrane on the pp-wave 
from several aspects. In Ref.\,\cite{SY2} we investigated 
BPS conditions of the supermembrane. The BPS states in the matrix model 
on the pp-wave have been intensively studied \cite{BPS}. 
It was also shown in Ref.\,\cite{Malda} that the giant graviton as 
a classical solution 
can exist due to the presence of the constant R-R flux \cite{Myers}. 
Then, in particular, we 
showed the quantum stability of the giant graviton (supersymmetric fuzzy
sphere solution) and instability of the non-supersymmetric fuzzy sphere
solution in Ref.\,\cite{SY3}. As well as by our work, the 
classical solutions in the matrix model have been widely investigated
\cite{bak}.  
Moreover, in Ref.\,\cite{SY4} 
we also derived the less supersymmetric
background with 24 supercharges from the KG solution and 
studied the type IIA string from the supergravity analysis and 
the double dimensional reduction \cite{DDR}. Furthermore, 
we also investigated matrix string theories.  
The supersymmetries of 
the above type IIA string were initially discussed in \cite{HS}. 

In this paper we will continue 
to study the supermembrane on the pp-wave and consider 
the ground-state  wave function of the 
supermembrane on the pp-wave by 
the use of the quantum-mechanical procedure of 
de Wit-Hoppe-Nicolai \cite{dWHN}. It is well-known that a single
supermembrane in the flat space is unstable \cite{unstable}. 
This instability is inherently based 
on the continuous spectrum of this system. 
However, the action of the supermembrane (or matrix model) 
on the pp-wave contains the bosonic and the 
fermionic mass terms and the Myers term, 
and hence it would be possible to 
expect that the spectrum of the system might be discrete 
or that a single supermembrane might be stabilized.  
Thus, it would be very interesting to study the ground state  or spectrum
of this system. To accomplish this, 
the complex-spinor notation ($Spin(7)\times U(1)$ formalism) 
will be introduced. We calculate the superalgebra of complex 
supercharges.  
Then, we will investigate the 
zero-mode Hamiltonian of the system, whose spectrum 
is non-trivial in contrast to the flat case.  
This system is identical with the abelian matrix model 
describing the motion of 
superparticles on the pp-wave. The Hamiltonian resembles the
supersymmetric harmonic oscillator but the supercharges do not commute
with the Hamiltonian. 
The supergravity multiplet ${\bf 128}$ (boson) 
+ ${\bf 128}$ (fermion) in the flat space comes to split 
non-trivially. As a result, a unique supersymmetric 
ground state  is left, and all the other states are lifted up.
Also, we show that the zero-mode Hamiltonian is positive definite and  
explicitly construct the ground-state wave function for the zero-mode 
Hamiltonian. It is represented by the product of the ground states  of the
bosonic harmonic oscillators and the fermionic variables. 
For this ground state, both the orbital and the spin angular momenta 
of $SO(3)\times SO(6)$ vanish.  
 
Furthermore, we investigate the nonzero-mode Hamiltonian 
and its ground-state wave function. For example, by truncating the
variables, the supersymmetric ground-state  wave function is explicitly 
constructed in the $SU(2)$ matrix model case. 
This ground state seems non-normalizable, 
but its $L^2$-norm can be represented by 
an asymptotic series with respect to $1/\mu^3$. The zeroth order term
corresponds to the kinematical contribution to the ground state, while 
the corrections with the powers of $1/\mu^3$ are related 
to the dynamical contributions 
arising from non-diagonal elements of matrix variables. 
In the IR region, 
the matrix variables are restricted to the abelian part. 
From these considerations, it might be understood 
that the dynamical effect tends to spoil
down the normalizability of the zero-mode ground state  as the system
flows from the IR region to the UV one, at least in our example.

This paper is organized as follows. 
In section 2 we 
introduce the complex-spinor notation by 
following the quantum-mechanical procedure of de Wit-Hoppe-Nicolai. 
Section 3 is devoted to the study of the zero-mode Hamiltonian of the
system. We can obtain the spectrum, and 
construct the unique supersymmetric ground state.   
Also, the positive definiteness of the zero-mode Hamiltonian is shown.  
In section 4 we discuss the nonzero-mode Hamiltonian and 
construct an example of the ground-state wave function with the 
truncation of the variables in $N=2$ case.  
Its $L^2$-norm can be represented by an asymptotic series.  
Section 5 is devoted to conclusions and
discussions. In appendix, we discuss that 
the ground state of the zero-mode Hamiltonian is an $SO(3)\times SO(6)$ singlet.

\section{Supercharges and Superalgebra}

In this section, we shall start with our setup and 
introduce the complex-spinor notation ($Spin(7)\times U(1)$ formalism), 
which is convenient to study the ground state in later sections.  
We will also obtain the complex representation of the 
superalgebra.

\subsection{Setup}

The Lagrangian of the supermembrane on the KG background in 
the light-cone gauge 
is written in terms of an $SO(9)$ spinor $\psi$ as 
\begin{eqnarray}
 w^{-1}\mathcal{L} &=& \frac{1}{2}D_{\tau}X^r D_{\tau}X^r 
- \frac{1}{4}(\{X^r,\,X^s\})^2 - \frac{1}{2}\left(\frac{\mu}{3}\right)^2
\sum_{{\s I}=1}^{3}X_{\s I}^2 - \frac{1}{2}\left(\frac{\mu}{6}\right)^2 
\sum_{{\s I'}=4}^{9}X^2_{\s I'}    \nn \\
& & - \frac{\mu}{6} \sum_{{\s I,J,K}=1}^3\epsilon_{\s IJK}X^{\s K}
\{X^{\s I},\,X^{\s J}\} + i {\psi}^{\s T}\gamma^{r}\{X^r,\,\psi \} 
+ i{\psi}^{\s T} D_{\tau}\psi + i\frac{\mu}{4}\psi^{\s T}\gamma_{\s 123} 
\psi \, .
\label{memb}
\end{eqnarray}
Here ``$\tau$'' is the time coordinate on the
worldvolume and 
$\{\, , \, \}$ is Lie bracket given using  
an arbitrary function $w(\sigma)$ of 
worldvolume spatial coordinates $\sigma^a$ ($a=1,2$)
\[
 \{A\,, B\} \; \equiv \; \frac{1}{w}
\epsilon^{ab} \partial_{a} A \partial_{b} B \, ,
\quad (\,a,b = 1,2\,) .
\]
with $\partial_a \equiv \frac{\partial}{\partial \sigma^a}$.
Also this theory has large residual gauge symmetry called 
the area-preserving diffeomorphism (APD) 
and the covariant derivative for this gauge symmetry
is defined by a gauge connection $\omega$ 
\begin{equation}
 D_{\tau} X^r \;\equiv\; \pd_{\tau}X^r - \{\omega,\,X^r \}\, .
\end{equation} 
The original symmetries cannot be manifestly seen in the light-cone gauge, 
but the Lagrangian (\ref{memb}) still has residual supersymmetries, 
\begin{eqnarray}
 & & \del_{\epsilon} X^r \;=\; 2\psi^{\s T}\gamma^r \epsilon(\tau)\, , 
\quad \del_{\epsilon} \omega \;=\; 2\psi^{\s T} \epsilon(\tau)\, , \nn \\
\label{linear}
& &  \del_{\epsilon}\psi \;=\; - i D_{\tau} X^r \gamma_r \epsilon(\tau) 
\,+\, \frac{i}{2}\{X^r,\, X^s \}\gamma_{rs}\epsilon(\tau)  \\
& & \,\qquad\qquad  + \, \frac{\mu}{3} i 
\sum_{I=1}^{3}X^{\s I} \gamma_{\s I}\gamma_{\s 123}\epsilon(\tau) 
\,-\, \frac{\mu}{6} i\sum_{I'=4}^{9} X^{\s I'}\gamma_{\s I'}
\gamma_{\s 123}
\epsilon(\tau)\, ,\nn \\
& & \epsilon(\tau) \;=\; \exp\left(\frac{\mu}{12}\gamma_{\s 123}\tau \right)
\epsilon_0\, \quad (\,\epsilon_0:\,{\rm constant~spinor}). \nn
\end{eqnarray}
These transformation rules are 16 linearly-realized supersymmetries on
the maximally supersymmetric pp-wave. 
In taking the limit $\mu \rightarrow 0$,  
we recover the supersymmetry   
transformations in the flat space. 
In the context of the eleven-dimensional supersymmetries, these
correspond 
to the dynamical supersymmetries.
The Lagrangian (\ref{memb}) has another 16 nonlinearly realized
supersymmetries,
\begin{eqnarray}
 & & \del_{\eta}X^r \;=\; 0\,,\quad \del_{\eta}\omega \;=\; 0\, , \nn \\
\label{nonlinear}
 & & \del_{\eta}\psi \;=\; \eta(\tau)\, ,  \\
 & & \eta(\tau) \;=\; \exp\left( - \frac{\mu}{4}\gamma_{\s 123} \tau \right)
\eta_0\, , \quad (\,\eta_0:\,{\rm constant~spinor}), \nn
\end{eqnarray} 
which correspond to the kinematical supersymmetries in the 
eleven-dimensional theory.

In our previous work \cite{SY}, we derived the 
supercharges and superalgebra of the supermembrane theory 
on the pp-wave background using the $SO(9)$ formalism.  
The supercharges $Q^+$ for the dynamical supersymmetries and $Q^-$ for
the kinematical supersymmetries are obtained as Noether charges 
\begin{eqnarray}
& & Q^+ = \int \! d^2\sigma\, w 
\Bigg[ -2 \e^{-\frac{\mu}{12}\gamma_{\s 123}\tau} 
\Big( DX^r \gamma_r\psi + \frac{1}{2}\{X^r,\,X^s\}\gamma_{rs}\psi \nn \\
& &\qquad  + \frac{\mu}{3}\sum_{\s I=1}^3
X^{\s I}\gamma_{\s I}\gamma_{\s 123}\psi 
+ \frac{\mu}{6}\sum_{\s I'=4}^9 X^{\s I'}\gamma_{\s I'}\gamma_{\s 123}\psi 
\Big)\Bigg]\,, 
\label{ch-l}
\\
&&Q^- = \int \! d^2\sigma\, 
w\left[-2i \e^{\frac{\mu}{4}\gamma_{\s 123}\tau}\psi
\right]\, \nn \\
&&\qquad = -2i \e^{\frac{\mu}{4}\gamma_{\s 123}\tau}\psi_0\, ,
\label{ch-nl}
\end{eqnarray}
where $\psi_0$ is the zero-mode of $\psi$ and we used the
normalization with
$\int \! d^2\sigma\, w(\sigma)= 1$. 

By the use of the Dirac brackets, 
\begin{eqnarray}
&&\{X^r,D_{\tau}X_s\}_{\rm\s DB}=\frac{1}{w}\delta_{s}^{r}\delta^{(2)}(\sigma-\sigma')\,,\label{DB-bo}\\
&& \{\psi_{\al}(\sigma),\,\psi^{\s T}_{\beta}(\sigma') \}_{\rm\s DB} = 
-\frac{i}{2w}\del_{\al\beta}\del^{(2)}
(\sigma - \sigma')\,,
\label{DB-fermion}
\end{eqnarray}
the superalgebra can be calculated 
and we obtain the following results 
\begin{eqnarray}
& & i\left\{\frac{1}{\sqrt{2}}Q_{\al}^-,\,\frac{1}{\sqrt{2}}
(Q^-)_{\beta}^{\s T}\right\}_{\rm\s DB} \;=\; 
 - \del_{\al\beta}\, , \\
& & i\left\{\frac{1}{\sqrt{2}}Q_{\al}^+,\,\frac{1}{\sqrt{2}}
(Q^-)_{\beta}^{\s T}\right\}_{\rm\s DB} \;=\; 
  i\sum_{{\s I}=1}^{3} \left[ \left(P_0^{\s I} + \frac{\mu}{3}
X^{\s I}_0 \gamma_{\s 123}\right)\gamma_{\s I} 
\e^{-\frac{\mu}{3}\gamma_{\s 123}\tau}
\right]_{\al\beta}  \\
& & \qquad  +  i\sum_{{\s I'}=4}^{9} \left[ \left(
P_0^{\s I'} - \frac{\mu}{6}X^{\s I'}_0
\gamma_{\s 123} \right)\gamma_{\s I'} 
\e^{-\frac{\mu}{6}\gamma_{\s 123}\tau}
\right]_{\al\beta}\, , \nn \\
& & i\left\{\frac{1}{\sqrt{2}}Q_{\al}^+,\,\frac{1}{\sqrt{2}}
(Q^+)_{\beta}^{\s T}\right\}_{\rm\s DB} \;=\; 
2H \delta_{\alpha\beta}  \\
& & \qquad +  \frac{\mu}{3}\sum_{{\s I,J}=1}^3 M^{\s IJ}_0 
\left(\gamma_{\s IJ}\gamma_{\s 123}\right)_{\al\beta} 
- \frac{\mu}{6}\sum_{{\s I',J'}=4}^9 
M_0^{\s I'J'}
\left(\gamma_{\s I'J'}\gamma_{\s 123}\right)_{\al\beta} \nn \\
& & \qquad - 2 \sum_{{\s I}=1}^3
\int\! d^2\sigma\, \varphi X_{\s I}(\gamma^{\s I})_{\al\beta}
-2\sum_{{\s I'=4}}^9\int\!d^2\sigma\,\varphi X_{\s I'}
\left(\gamma^{{\s I'}}\e^{\frac{\mu}{6}\gamma_{\s 123}\tau}\right)_{\al\beta}
\,, \nn 
\end{eqnarray}
where we omitted the central charges since 
these are not discussed in this paper.   
Here $\varphi$ describes the constraint condition, 
\begin{eqnarray}
\varphi \;=\; w\{w^{-1}P^r,\,X^r\} + iw\{\psi^{\s T},\,\psi\}\,.
\end{eqnarray}
The $P^r \equiv wD_{\tau}X^r$ and $S_{\al} = iw\psi^{\s T}_{\al}$ 
are canonical momenta of the $X^r$ and $\psi$, respectively. 
The $M^{\s IJ}$ and $M^{\s I'J'}$ are defined by 
\begin{eqnarray}
& &  M^{\s IJ} \;=\; X^{\s I}P^{\s J} - P^{\s I}X^{\s J} - 
\frac{1}{2}S \gamma^{\s IJ}\psi\,, \\
& &  M^{\s I'J'} \;=\; X^{\s I'}P^{\s J'} - P^{\s I'}X^{\s J'} - 
\frac{1}{2}S \gamma^{\s I'J'}\psi\,, 
\end{eqnarray} 
and the $SO(3)\times SO(6)$ Lorentz generators 
$M^{\s IJ}_0$ and $M^{\s I'J'}_0$ are given as
\begin{eqnarray}
& & M^{\s IJ}_0 \;\equiv\; \int\!d^2\sigma\,M^{\s IJ}\, , \\
& & M^{\s I'J'}_0 \;\equiv\; \int\!d^2\sigma\,M^{\s I'J'}\, .
\end{eqnarray}
They satisfy the $SO(3)\times SO(6)$ Lorentz algebra, 
\begin{eqnarray}
& & \left\{M_0^{\s IJ},\, M_0^{\s KL} \right\}_{\rm DB} \;=\;
\del^{IK}M_0^{\s JL} - \del^{\s IL}M_0^{\s JK} - \del^{\s JK}M_0^{\s IL} 
+ \del^{\s JL}M_0^{\s IK}\,, \\
& & \left\{M_0^{\s I'J'},\, M_0^{\s K'L'} \right\}_{\rm DB} \;=\;
\del^{I'K'}M_0^{\s J'L'} - \del^{\s I'L'}M_0^{\s J'K'} - \del^{\s J'K'}
M_0^{\s I'L'} 
+ \del^{\s J'L'}M_0^{\s I'K'}\,.
\end{eqnarray}
The zero-modes of $P^r (\equiv w D_{\tau}X^{r})$ and $X^r$ are 
written by 
\begin{eqnarray}
 P_0^r &\equiv& \int\! d^2\sigma\, w D_{\tau}X^{r}\, , 
\,\,\,X^{r}_0 \;\equiv\; \int\!d^2\sigma\,wX^{r}\,. 
\end{eqnarray}
Also, the Hamiltonian $H$ is given by 
\begin{eqnarray}
& & H \;=\; \int\!d^2\sigma\,w\Bigg[\frac{1}{2}\left(\frac{P^r}{w}\right)^2
+ \frac{1}{4} \{X^r,X^s\}^2
+ \frac{1}{2} \left(\frac{\mu}{3}\right)^2\sum_{{\s I}=1}^{3}(X^{\s I})^2
+ \frac{1}{2} \left(\frac{\mu}{6}\right)^2\sum_{{\s I'}=4}^{9}(X^{\s I'})^2
\nn\\
& &\qquad +\frac{\mu}{6}\sum_{{\s I,J,K}=1}^{3}\epsilon_{\s IJK}
X^{\s K} \{X^{\s I},X^{\s J} \} - w^{-1}\frac{\mu}{4}S\gamma_{\s 123}
\psi -w^{-1}S\gamma_{r}\{X^r,\psi\}\Bigg]\,.
\end{eqnarray}
In the next section, we will introduce the 
complex spinor notation in order to study the 
ground-state wave function.

\subsection{$Spin(7)\times U(1)$ Formalism}

Here we shall introduce the $Spin(7) \times U(1)$ formalism 
following Ref.\,\cite{dWHN}. 
Hereafter, we use the decomposition of the $SO(9)$ 
gamma matrices into the $SO(7)$ ones as follows: 
\begin{eqnarray}
 \gamma_i &=& \Gamma_i \otimes \sigma_2 \;=\; 
\begin{pmatrix}
0 & -i\Gamma_i \\
i\Gamma_i & 0 
\end{pmatrix}
\,, \quad (i=1,\,\ldots,\,7)\,,  \\
 \gamma_8 &=& {\bf 1}_8 \otimes \sigma_1 \;=\; 
\begin{pmatrix}
0 & {\bf 1}_8 \\
{\bf 1}_8 & 0 
\end{pmatrix}
\,, \quad \gamma_9 \;=\; {\bf 1}_8 \otimes \sigma_3 \;=\; 
\begin{pmatrix}
{\bf 1}_8 & 0 \\
0 & - {\bf 1}_8 
\end{pmatrix}
\,. \nn 
\end{eqnarray}
The above $SO(7)$ gamma matrices satisfy the following relations:
\begin{eqnarray}
 (\Gamma_i)^{\dagger} \;=\; \Gamma_i\,, \quad (\Gamma_i)^{\s T} \;=\; 
- \Gamma_i\,, \quad (\Gamma_i)^{\ast} \;=\; - \Gamma_i\,.
\end{eqnarray}

To begin, let us decompose the real 16-component $SO(9)$ spinor 
into two real 8-component spinors 
$\psi^{(+)}$ and $\psi^{(-)}$, 
which are eigenspinors for the chirality matrix $\gamma_9$ 
defined by 
\begin{eqnarray}
 \gamma_9 \psi^{(\pm)} \;=\; (\pm 1) \cdot \psi^{(\pm)}\,.
\end{eqnarray}
In this time we can express eigenspinors as 
$\psi^{(+)}_{\al} = \psi_{\al}$ and $\psi^{(-)}_{\beta} = \psi_{8+\beta}
~(\al,\,\beta = 1,\,\ldots,\,8)$. 
Here, we shall introduce a complex spinor $\lambda$ 
defined by 
\begin{eqnarray}
 \lambda_{\al} \;=\; \psi_{\al}^{(+)} + i \psi_{\al}^{(-)}\,, \quad 
(\al =1,\,\ldots,\,8)\,,
\end{eqnarray} 
that is, we express  the real 16-component $SO(9)$ spinor 
as a single 8-component complex spinor $\lambda$. This complex 
spinor is linearly-realized under the $SO(7)\times U(1)$ transformation,
which is the subgroup of the original $SO(9)$. The following
expression 
\begin{eqnarray}
 \psi &=& \dbinom{\psi^{(+)}}{\psi^{(-)}} \;=\; 
\frac{1}{2}\dbinom{\lambda + \lambda^{\dagger}}{-i(\lambda 
- \lambda^{\dagger})}
\end{eqnarray}
is also useful for rewriting the supercharges. 

The bosonic coordinates $X^8$ and $X^9$ are combined into a complex 
coordinate 
\begin{eqnarray}
 Z &=& \frac{1}{\sqrt{2}} (X^8 + iX^9)\,, 
\end{eqnarray}
which transforms linearly under the $U(1)$ subgroup. 
The canonical momentum for $Z$ is given by 
\begin{eqnarray}
 \mathcal{P} &=& \frac{1}{\sqrt{2}}(P^8 - i P^9)\,. 
\end{eqnarray} 
By the standard procedure, the Dirac brackets 
for these variables 
are given by 
\begin{eqnarray}
\{X^i(\tau,\sigma),\,P^j(\tau,\sigma')\}_{\rm\s DB} &=& \del^{ij}
\del^{(2)}(\sigma - \sigma')\,,\quad (i,j=1,\,\ldots,\,7)\,, \nn \\
\{Z(\tau,\sigma),\,\mathcal{P}(\tau,\sigma')\}_{\rm\s DB} 
&=& \del^{(2)}(\sigma - \sigma')\,,  \\
\{\lambda_{\al}(\tau,\sigma),\,
\lambda_{\beta}^{\dagger}(\tau,\sigma')\}_{\rm\s DB} 
&=& - \frac{i}{w}\del_{\al\beta}\del^{(2)}(\sigma - \sigma')\,, \quad 
(\al,\beta = 1,\,\ldots,\,8). \nn
\end{eqnarray}

Next, let us rewrite the supercharge $Q^+$ in terms of the complex spinor
$\lambda$. We decompose the supercharge $Q^+$ into $(Q^{+})^{(\pm)}$ 
as 
\[
 Q^+ \;=\; \dbinom{(Q^+)^{(+)}}{(Q^+)^{(-)}}\,, 
\]
and define the complex supercharge $Q$ by 
\begin{eqnarray}
Q &\equiv& \frac{1}{2}\left( (Q^+)^{(+)} +i (Q^+)^{(-)} \right)\,. 
\end{eqnarray}
Thus, after straightforward calculation, we can obtain 
the expression of the complex supercharge as 
\begin{eqnarray}
Q &=& \int\!d^2\sigma\, \e^{\frac{\mu}{12}\Gamma_{123}\tau}
\Biggl[\Bigl(\, P^i\Gamma_i - \frac{1}{2}w\{X^i,\,X^j\}\Gamma_{ij} 
+ w\{Z,\,\bar{Z}\} \nn \\
&& \quad - \frac{\mu}{3}w\sum_{{\s I}=1}^3 
X^{\s I}\Gamma_{\s I}\Gamma_{123} - \frac{\mu}{6}w \sum_{m=4}^7 
X^m\Gamma_m\Gamma_{123}\Bigr)\lambda \nn \\
&& \quad - \sqrt{2}\,i\Bigl(
\mathcal{P} - w\{X^i,\,\bar{Z}\}\Gamma_i + \frac{\mu}{6}w\bar{Z}\Gamma_{123}
\Bigr)\lambda^{\dagger}\,\, 
\Biggr]\,,
\end{eqnarray} 
and its complex conjugate is given by 
\begin{eqnarray}
Q^{\dagger} &=& -\int\!d^2\sigma\, \e^{-\frac{\mu}{12}\Gamma_{123}\tau}
\Biggl[\Bigl(\, P^i\Gamma_i + \frac{1}{2}w\{X^i,\,X^j\}\Gamma_{ij} 
+ w\{Z,\,\bar{Z}\} \nn \\
&& \quad + \frac{\mu}{3}w\sum_{{\s I}=1}^3 
X^{\s I}\Gamma_{\s I}\Gamma_{123} + \frac{\mu}{6}w \sum_{m=4}^7 
X^m\Gamma_m\Gamma_{123}\Bigr)\lambda^{\dagger} \nn \\
&& \quad - \sqrt{2}\,i\Bigl(
\bar{\mathcal{P}} 
+ w\{X^i,\,Z\}\Gamma_i - \frac{\mu}{6}w Z\Gamma_{123}
\Bigr)\lambda\,\, 
\Biggr]\,.
\end{eqnarray} 
The superalgebra of these complex supercharges is computed as follows: 
\begin{eqnarray}
i\{ Q_\alpha , Q_\beta \}_{\rm\s DB}
&=& i\frac{\mu}{3} \sum_{m=4}^{7} M_0^{m+}
\left( \Gamma_m \Gamma_{123} \right)_{\alpha \beta}
-2 \sqrt{2}i \int\! d^2 \sigma\, 
\varphi \bar{Z} \left( e^{\frac{\mu}{6} 
\Gamma_{123}\tau} \right)_{\alpha \beta}\,,
\\
i\{ Q_\alpha^\dagger , Q_\beta^\dagger \}_{\rm\s DB}
&=& - i\frac{\mu}{3} \sum_{m=4}^{7} M_0^{m-}
\left( \Gamma_m \Gamma_{123} \right)_{\alpha \beta}
+2 \sqrt{2}i \int \!d^2 \sigma\, \varphi Z \left( e^{-\frac{\mu}{6} 
\Gamma_{123}\tau} \right)_{\alpha \beta}\,, \\
i\{ Q_\alpha , Q_\beta^\dagger \}_{\rm\s DB}
&=& 2 H \delta_{\alpha \beta}
+ i\frac{\mu}{3}M_0^{89} \left( \Gamma_{123} \right)_{\alpha \beta}
- \frac{\mu}{3} \sum_{I,J=1}^{3} M_{0}^{IJ} 
\left( \Gamma_{IJ} \Gamma_{123} \right)_{\alpha \beta} \\
&& + \frac{\mu}{6} \sum_{m,n=4}^{7} M_{0}^{mn}
\left( \Gamma_{mn} \Gamma_{123} \right)_{\alpha \beta}
+ 2 \sum_{I=1}^{3} \int\!d^2 \sigma\, \varphi X_{I} 
\left( \Gamma_I \right)_{\alpha \beta}
\nonumber \\
&& + 2 \sum_{m=4}^{7} \int\! d^2 \sigma\, \varphi X_{m} 
\left( \Gamma_m e^{-\frac{\mu}{6}\Gamma_{123}\tau}
 \right)_{\alpha \beta}\,, \nn 
\label{algebla}
\end{eqnarray}
where the zero-modes of angular momenta $M_0^{\s IJ}$, $M_0^{mn}$, 
$M_0^{m\pm}$ and $M_0^{89}$ are rewritten in terms of the complex
notation as follows:   
\begin{eqnarray}
M_0^{\s IJ} &=& \int\!d^2\sigma\,M^{\s IJ}\,, \quad 
M^{\s IJ} \;=\; X^{\s I}P^{\s J} - X^{\s J}P^{\s I} 
- iw \frac{1}{2}\lambda^{\dagger}\Gamma^{\s IJ}\lambda\,, \\  
M_0^{mn} &=& \int\!d^2\sigma\,M^{mn}\,, \quad 
M^{mn} \;=\; X^{m}P^{n} - X^{n}P^{m} 
- iw \frac{1}{2}\lambda^{\dagger}\Gamma^{mn}\lambda\,, \\  
M_0^{m+} &=& \int\! d^2 \sigma\, M^{m+}\,, \quad 
M^{m+} \;=\; \sqrt{2} \left( X^m {\cal P} - P^m \bar{Z} \right)
- \frac12 w \lambda \Gamma_m \lambda\,,
\\
M_0^{m-} &=& \int\! d^2 \sigma\, M^{m-}\,, \quad
M^{m-} \;=\; \sqrt{2} \left( X^m \bar{\cal P} - P^m Z \right)
- \frac12 w \lambda^\dagger  \Gamma_m \lambda^\dagger\,, \\ 
M_0^{89} &=& \int\!d^2\sigma\,M^{89}\,, \quad    
M^{89} \;=\; i \left( Z {\cal P} - \bar{Z} \bar{\cal P} \right)
+ \frac{1}{4}w\left(\lambda^{\dagger}\lambda 
- \lambda\lambda^{\dagger}\right)\,.
\end{eqnarray}
In the above superalgebra, the Hamiltonian $H$ is given by 
\begin{eqnarray}
H &=& \int \!d^2\sigma\, w\Biggl[\frac{1}{2}\left(\frac{P^i}{w}\right)^2 
+ w^{-2}|\mathcal{P}|^2 + \frac{1}{4}\left\{X^i,\,X^j\right\}^2 
+ |\{X^i,\,Z\}|^2 + \frac{1}{2}|\{Z,\,\bar{Z}\}|^2 \nn \\
&& + \frac{1}{2}\left(\frac{\mu}{3}\right)^2\sum_{{\s I}=1}^3(X^{\s I})^2 
+ \frac{1}{2}\left(\frac{\mu}{6}\right)^2\sum_{m=4}^7(X^{m})^2 
+ \left(\frac{\mu}{6}\right)^2 Z\bar{Z} 
+ \frac{\mu}{6}\sum_{{\s I,J,K}=1}^3\epsilon_{\s IJK}X^{\s K}\{X^{\s I},\,
X^{\s J}\} \nn \\ 
&& -i\lambda\Gamma_i\{X^i,\,\lambda^{\dagger}\} - \frac{1}{\sqrt{2}}\left(
\lambda\{Z,\,\lambda\} - \lambda^{\dagger}\{\bar{Z},\,\lambda^{\dagger}\}
\right)
- i\frac{\mu}{4}\lambda\Gamma_{123}\lambda^{\dagger}
\Biggr]\,.
\end{eqnarray}
The constraint condition is also expressed by 
\begin{eqnarray}
w^{-1}\varphi \;=\; \{w^{-1}P^i, X^i \}+\{ w^{-1}{\cal P}, Z \}
+\{ w^{-1}\bar{\cal P},\bar{Z} \}
+i \{\lambda, \lambda^\dagger  \} \,.
\end{eqnarray}
When we contract the spinor indices $\al$ and $\beta$, 
the Hamiltonian is represented with the above supercharges as 
\begin{eqnarray}
H &=& \frac{1}{16}i\{Q_{\al},Q^{\dagger}_{\al}\}_{\rm\s DB}\,.
\end{eqnarray}

Now, we shall move to a matrix representation by replacing the 
variables with matrices, according to: 
\begin{eqnarray}
 X(\xi^{i}) & \longrightarrow & X(\tau)\,, \nn \\
 \psi(\xi^i) & \longrightarrow & \psi(\tau)\,, \nn \\
 \int\!d^2\sigma\, w(\sigma) & \longrightarrow & N{\rm Tr}\,, \nn \\
 \{~ , ~\} & \longrightarrow & -i[~ , ~ ]\,. \nn  
\end{eqnarray}
In this case, 
the Hamiltonian is rewritten into 
 \begin{eqnarray}
H &=& N\mbox{Tr}\Biggl[\frac12 \left(P^i \right)^2 
+ \mathcal{P} \bar{\mathcal{P}} 
- \frac14 \left[X^i,X^j \right]^2
+ \left| \left[ X^i, Z \right] \right|^2
+ \frac12 \left| \left[ Z, \bar{Z} \right] \right|^2
\nonumber \\
&&+ \frac12 \left( \frac{\mu}{3}\right)^2 \sum_{{\s I}=1}^{3} 
\left( X^{\s I} \right)^2
+ \frac12 \left( \frac{\mu}{6}\right)^2 \sum_{m=4}^{7} \left( X^m \right)^2
+ \left( \frac{\mu}{6} \right)^2 Z \bar{Z}
-i \frac{\mu}{3}\sum_{{\s I,J,K}=1}^{3} 
\epsilon_{\s IJK}X^{\s I}X^{\s J}X^{\s K}
\nn \\
&&- \lambda \Gamma_{i} \left[X^i , \lambda^\dagger \right]
+ \frac{1}{\sqrt2}i \left( 
\lambda \left[Z, \lambda\right] -\lambda^\dagger \left[\bar{Z}, 
\lambda^\dagger\right]
\right)
- i \frac{\mu}{4} \lambda \Gamma_{123} \lambda^\dagger
\Biggr],
\end{eqnarray}
where the canonical momenta are redefined by 
$P^i \equiv D_{\tau}X^i,~
{\cal P} \equiv D_{\tau}\bar{Z},~
\bar{\cal P} \equiv  D_{\tau}Z$. 
The variables are expressed by the $U(N)$
generators $T^A~(A=0,\,1,\,\ldots,\,N^2 -1,~T^0={\bf 1})$ as
follows:\footnote{We 
assume that the topology of the membrane is sphere.}
\begin{eqnarray}
X^i &=& X_0^i {\bf 1} + \sum_{A=1}^{N^2-1}X_A^iT^A\,, 
\quad 
P^i \;=\; P_0^i {\bf 1} + \sum_{A=1}^{N^2-1}P_A^iT^A\,, \\
Z &=& Z_0 {\bf 1} + \sum_{A=1}^{N^2-1}Z_AT^A\,, 
\quad \mathcal{P} \;=\; \mathcal{P}_0{\bf 1} + \sum_{A=1}^{N^2-1}
\mathcal{P}_AT^A\,, \\
\lambda_{\al} &=& \lambda_{0\al}{\bf 1} 
+ \sum_{A=1}^{N^2 -1}\lambda_{A\al}T^A\,, \quad 
\lambda^{\dagger}_{\al} \;=\; \lambda_{0\al}^{\dagger}{\bf 1} + 
\sum_{A=1}^{N^2-1}\lambda_{A\al}^{\dagger}T^A
\,.
\end{eqnarray}
Also, the $U(N)$ generators $T^{A}~(A=0,1,\,\ldots,\,N^2-1)$ 
satisfy the following relations: 
\[
 [T^{A},\,T^{B}] \;=\; if_{ABC}T^{C}\,,\quad {\rm Tr}(T^AT^B)
\;=\; \frac{1}{N}\del^{AB}\,.
\]
where the structure constant $f_{ABC}$ is a 
completely antisymmetric tensor satisfying 
$f_{0AB}=0$. 

Now, the Dirac brackets are replaced with the following 
(anti-)commutation relations
\begin{eqnarray}
[X^i_{A},\,P^j_{B}] &=& i\del^{ij}\del_{AB}
\,,\quad (i,j=1,\,\ldots,\,7)\,, \nn \\ 
\left[ Z_{A},\,\mathcal{P}_{B}\right]  
&=& [\bar{Z}_{A},\,\bar{\mathcal{P}}_{B}] \;=\; 
i\del_{AB}\,,  \\
\left\{\lambda_{A \al},\,
\lambda_{B\beta}^{\dagger}\right\} 
&=& \del_{\al\beta}\del_{AB}\,, \quad 
(\al,\beta = 1,\,\ldots,\,8). \nn
\end{eqnarray}
Therefore, we can represent the conjugate momenta by differential 
operators 
\begin{eqnarray}
&& P_{A}^i \;=\; -i \frac{\partial}{\partial X^i_{A}}\,, \quad 
\mathcal{P}^{A} \;=\; -i \frac{\partial}{\partial Z_{A}}\,, \nn \\
&&\bar{\mathcal{P}}^A \;=\; -i \frac{\partial}{\partial\bar{Z}_{A}}\,, \quad 
\lambda^{\dagger}_{A\al} \;=\; \frac{\partial}{\partial\lambda_{A\al}}\,.
\end{eqnarray}
Then we obtain the following expressions for supercharges as 
\begin{eqnarray}
Q_{\al} &=& \left(\e^{\frac{\mu}{12}\Gamma_{123}\tau}\right)_{\al\beta}
\Biggl[\Bigl(\, -i(\Gamma_i)_{\beta\gamma}\frac{\partial}{\partial X_{A}^i} 
- \frac{1}{2}f_{ABC}X_{B}^i X_{C}^j(\Gamma_{ij})_{\beta\gamma} 
+ f_{ABC}Z^{B}\bar{Z}^{C}\del_{\beta\gamma}\nn \\
&& \quad - \frac{\mu}{3}\sum_{{\s I}=1}^3 
X^{\s I}_{A}(\Gamma_{\s I}\Gamma_{123})_{\beta\gamma} 
- \frac{\mu}{6}\sum_{m=4}^7 X^m_{A}
(\Gamma_m\Gamma_{123})_{\beta\gamma}\Bigr)\lambda_{A\gamma} \nn \\
&& \quad - \sqrt{2}\Bigl(
\frac{\partial}{\partial Z_{A}}\del_{\beta\gamma} 
- if_{ABC}X_{B}^i\bar{Z}_{C}(\Gamma_i)_{\beta\gamma} 
+ i\frac{\mu}{6}\bar{Z}_{A}(\Gamma_{123})_{\beta\gamma}
\Bigr)\frac{\partial}{\partial \lambda_{A\gamma}}\,\, 
\Biggr]\,,
\end{eqnarray} 
and its complex conjugate is given by 
\begin{eqnarray}
Q^{\dagger}_{\al} &=& - \left(\e^{-\frac{\mu}{12}\Gamma_{123}\tau}
\right)_{\al\beta}
\Biggl[\Bigl(\, -i(\Gamma_i)_{\beta\gamma}
\frac{\partial}{\partial X^i_{A}} 
+ \frac{1}{2}f_{ABC}X^i_{B}X^j_{C}(\Gamma_{ij})_{\beta\gamma} 
+ f_{ABC}Z_{B}\bar{Z}_{C}\del_{\beta\gamma} \nn \\
&& \quad + \frac{\mu}{3}\sum_{{\s I}=1}^3 
X^{\s I}_{A}(\Gamma_{\s I}\Gamma_{123})_{\beta\gamma} 
+ \frac{\mu}{6} \sum_{m=4}^7 
X^m_{A}(\Gamma_m\Gamma_{123})_{\beta\gamma}
\Bigr)\frac{\partial}{\partial \lambda_{A\gamma}}
\nn \\
&& \quad - \sqrt{2}\Bigl(
\frac{\partial}{\partial \bar{Z}_A}\del_{\beta\gamma}
+ if_{ABC}X^i_{B}Z_{C}(\Gamma_i)_{\beta\gamma} 
- i\frac{\mu}{6} Z_{A}(\Gamma_{123})_{\beta\gamma}
\Bigr)\lambda_{A\gamma}\,\, 
\Biggr]\,.
\end{eqnarray} 
In this representation, the superalgebra becomes 
\begin{eqnarray}
\{ Q_\alpha , Q_\beta \}
&=& i\frac{\mu}{3} \sum_{m=4}^{7} M_0^{m+}
\left( \Gamma_m \Gamma_{123} \right)_{\alpha \beta}
-2 \sqrt{2}i \varphi_{A} \bar{Z}_A 
\left( e^{\frac{\mu}{6}\Gamma_{123}\tau} \right)_{\alpha \beta}\,,
\\
\{ Q_\alpha^\dagger , Q_\beta^\dagger \}
&=& -i\frac{\mu}{3} \sum_{m=4}^{7} M_0^{m-}
\left( \Gamma_m \Gamma_{123} \right)_{\alpha \beta}
+2 \sqrt{2}i\varphi_{A} Z_A \left( e^{-\frac{\mu}{6} \Gamma_{123}\tau}
\right)_{\alpha \beta},, \\
\{ Q_\alpha , Q_\beta^\dagger \}
&=& 2 H \delta_{\alpha \beta}
- \frac{\mu}{3} \sum_{I,J=1}^{3} M_{0}^{IJ} 
\left( \Gamma_{IJ} \Gamma_{123} \right)_{\alpha \beta}
+ \frac{\mu}{6} \sum_{m,n=4}^{7} M_{0}^{mn}\left( \Gamma_{mn} \Gamma_{123} 
\right)_{\alpha \beta} 
\\
&& + i\frac{\mu}{3}M_0^{89} \left( \Gamma_{123} \right)_{\alpha \beta}
+ 2 \sum_{I=1}^{3}\varphi_{A} X^{I}_{A} \left( \Gamma_I \right)
_{\alpha \beta} + 2 \sum_{m=4}^{7} \varphi_{A} X^{m}_{A} 
\left( \Gamma_m e^{-\frac{\mu}{6} \Gamma_{123}\tau} 
 \right)_{\alpha \beta}\,, \nn
\end{eqnarray}
where the zero-modes of angular momenta are represented by
\begin{eqnarray}
M_0^{\s IJ} &=& -iX^{\s I}_A\frac{\partial}{\partial X^{\s J}_A}  
+ i X^{\s J}_{A}\frac{\partial}{\partial X^{\s I}_{A}}
- \frac{i}{2} \lambda_A \Gamma^{\s IJ}\frac{\partial}{\partial\lambda_A}\,, 
\label{am1} \\
M_0^{mn} &=& -iX^m_A\frac{\partial}{\partial X^{n}_{A}} + iX^n_A\frac{\partial}
{\partial X^{m}_{A}}  
- \frac{i}{2} \lambda_A \Gamma^{mn} \frac{\partial}{\partial\lambda_A}\,, 
\label{am2} \\
M_0^{m+} &=& -\sqrt{2}i \left( X^m_A \frac{\partial}{\partial Z_A} 
- \bar{Z}_A\frac{\partial}{\partial X_A^m} \right)
- \frac12 \lambda_A \Gamma^m \lambda_A\,, \label{am3} \\
M_0^{m-} &=& -\sqrt{2}i \left( X^m_A \frac{\partial}{\partial \bar{Z}_A} 
- \frac{\partial}{\partial X^{m}_A} Z_A \right)
- \frac12 \frac{\partial}{\partial\lambda_A}  \Gamma^m 
\frac{\partial}{\partial \lambda_A}
\,, \label{am4}
\\ 
M_0^{89} &=& Z_A \frac{\partial}{\partial Z_A} 
- \bar{Z}_A \frac{\partial}{\partial \bar{Z}_A} 
- \frac12  \lambda_A\frac{\partial}{\partial \lambda_A} 
+ 2  \cdot {\rm dim}U(N)\,. 
\label{am5}
\end{eqnarray}
and the constraint condition $\varphi_A$ is expressed by 
\begin{eqnarray}
\varphi_{A} \;\equiv\; if_{ABC}\left(
X_B^i\frac{\partial}{\partial X_C^i} 
+ Z_B\frac{\partial}{\partial Z_C} + 
\bar{Z}_B \frac{\partial}{\partial \bar{Z}_C}
+ \lambda_B \frac{\partial}{\partial\lambda_C} \right)
\,. 
\label{constraint}
\end{eqnarray} 
It is assumed that physical states should satisfy the constraint
condition i.e., $\varphi \Psi_{\rm phys} =0$. 
Our purpose is to study the ground state. Thus, 
we shall also restrict ourselves to the $SO(3) \times SO(6)$ singlet 
satisfying 
\begin{equation}
M^{IJ}\Psi_{\rm phys}=M^{mn}\Psi_{\rm phys} = M^{m\pm}\Psi_{\rm phys} 
= M^{89}\Psi_{\rm phys} = 0\,,
\end{equation} 
since the ground state should have no angular momenta.   
The discussion on the states with nonzero angular momenta
would be also possible, but we do not consider them in this paper.

Furthermore, the associated Hamiltonian is written down as 
\begin{eqnarray}
 H &=& \frac{1}{16}\{Q_{\al},\,Q^{\dagger}_{\al}\} \nn \\
   &=& - \frac{1}{2}\frac{\partial^2}{\partial X^{i}_A \partial X^i_A}  
- \frac{\partial^2}{\partial Z_{A} \partial \bar{Z}_{A}} 
+ \frac{1}{4}f_{ABC}f_{ADE}X_{B}^{i}X_{C}^{j}X^{i}_{D}X^j_{E}  \\
&& + f_{ABC}f_{ADE}X^i_{B}Z_{C}X^i_{D}\bar{Z}_{E}
+ \frac{1}{2}f_{ABC}f_{ADE}Z_{B}\bar{Z}_{C}\bar{Z}_{D}Z_{E}
 \nn \\
&& + \frac{1}{2}\left(\frac{\mu}{3}\right)^2\sum_{{\s I}=1}^3(X^{\s I}_A)^2 
+ \frac{1}{2}\left(\frac{\mu}{6}\right)^2\sum_{m=4}^7(X^{m}_A)^2 
+ \left(\frac{\mu}{6}\right)^2 Z_{A}\bar{Z}_{A} 
+ \frac{\mu}{6}\sum_{{\s I,J,K}=1}^3\epsilon_{\s IJK}f_{ABC}
X_{A}^{\s I}X_{B}^{\s J}X_{C}^{\s K} \nn \\
&& + if_{ABC}X^i_{A}\lambda_{B}\Gamma_i\frac{\partial}{\partial \lambda_{C}}
+ \frac{1}{\sqrt{2}}f_{ABC}\left(
Z_{A}\lambda_{B}\lambda_{C} - \bar{Z}_{A}
\frac{\partial}{\partial\lambda_{B}}\frac{\partial}{\partial\lambda_{C}}
\right)
- i\frac{\mu}{4}\lambda_{A}\Gamma_{123}\frac{\partial}{\partial\lambda_{A}}
\,. \nn
\end{eqnarray}
Hereafter, by the use of the supercharges, the Hamiltonian, the angular momenta
and the constraint condition, we will study the ground-state wave function 
analytically.

\section{Supersymmetric Quantum Mechanics of Zero-mode Hamiltonian}

In this section, we shall consider the zero-mode ground-state wave function. 
We can extract the zero-mode part of the dynamical supercharges and 
the results are given by 
\begin{eqnarray}
Q_{\al}^{(0)} &=& \left(\e^{\frac{\mu}{12}\Gamma_{123}\tau}\right)_{\al\beta}
\Biggl[\Bigl(\, -i(\Gamma_i)_{\beta\gamma}\frac{\partial}{\partial X_{0}^i} 
- \frac{\mu}{3}\sum_{{\s I}=1}^3 
X^{\s I}_{0}(\Gamma_{\s I}\Gamma_{123})_{\beta\gamma} \nn \\
&& \quad - \frac{\mu}{6}\sum_{m=4}^7 X^m_{0}
(\Gamma_m\Gamma_{123})_{\beta\gamma}\Bigr)\lambda_{0\gamma} 
- \sqrt{2}\Bigl(
\frac{\partial}{\partial Z_{0}}\del_{\beta\gamma} 
+ i\frac{\mu}{6}\bar{Z}_{0}(\Gamma_{123})_{\beta\gamma}
\Bigr)\frac{\partial}{\partial \lambda_{0\gamma}}\,\, 
\Biggr]\,,
\label{ch1}
\end{eqnarray} 
and 
\begin{eqnarray}
Q^{(0)\dagger}_{\al} &=& - \left(\e^{-\frac{\mu}{12}\Gamma_{123}\tau}
\right)_{\al\beta}
\Biggl[\Bigl(\, -i(\Gamma_i)_{\beta\gamma}
\frac{\partial}{\partial X^i_{0}}
+ \frac{\mu}{3}\sum_{{\s I}=1}^3 
X^{\s I}_{0}(\Gamma_{\s I}\Gamma_{123})_{\beta\gamma} \nn \\
&& \quad + \frac{\mu}{6} \sum_{m=4}^7 
X^m_{0}(\Gamma_m\Gamma_{123})_{\beta\gamma}
\Bigr)\frac{\partial}{\partial \lambda_{0\gamma}}
- \sqrt{2}\Bigl(
\frac{\partial}{\partial \bar{Z}_{0}}\del_{\beta\gamma}
- i\frac{\mu}{6} Z_{0}(\Gamma_{123})_{\beta\gamma}
\Bigr)\lambda_{0\gamma}\,\, 
\Biggr]\,.
\label{ch2}
\end{eqnarray} 
The zero-mode Hamiltonian $H^{(0)}$ is obtained as 
\begin{eqnarray}
 H^{(0)} &=& \frac{1}{16}\{Q^{(0)}_{\al},\,Q^{(0)\dagger}_{\al}\}  \\ 
&=& H^{(0)}_{\rm B} + H^{(0)}_{\rm F}\,, \nn
\end{eqnarray}
where we defined the bosonic and fermionic Hamiltonians  
$H^{(0)}_{\rm B}$ and $H^{(0)}_{\rm F}$ by 
\begin{eqnarray}
H^{(0)}_{\rm B} &\equiv& 
- \frac{1}{2}\frac{\partial^2}{\partial X_0^i\partial X_0^i} 
- \frac{\partial^2}{\partial Z_0 \partial \bar{Z}_0} \nn \\
&& + \frac{1}{2}\left(\frac{\mu}{3}\right)^2\sum_{{\s I}=1}^3(X_0^{\s I})^2 
+ \frac{1}{2}\left(\frac{\mu}{6}\right)^2\sum_{m=4}^7(X_0^m)^2 
+ \left(\frac{\mu}{6}\right)^2 |Z_0|^2\,,  
\label{b-ham}
\\
H^{(0)}_{\rm F} 
&\equiv& - i\frac{\mu}{4}\lambda_{0\al}(\Gamma_{123})_{\al\beta}
\frac{\partial}{\partial\lambda_{0\beta}}
\,. \label{f-ham}
\end{eqnarray}
It is noted that the fermionic Hamiltonian for 
zero-modes is non-zero in the pp-wave case. Therefore, 
the nature of the supergravity multiplet is not simple and 
contains non-trivial results. The system described by the above 
zero-mode Hamiltonian resembles 
supersymmetric harmonic oscillator, although 
the supercharges do not commute with the Hamiltonian\footnote{
This system has already appeared in Ref.\,\cite{Malda} as the Lagrangian
for superparticles.}. 

Here, let us study the spectrum of this system quantum mechanically. 
To begin, we decompose the wave function as 
$\Psi^{(0)}_{\rm B}\otimes\Psi^{(0)}_{\rm F}$. 
We would like to study the 
states of the system. We have to study the energy 
eigenvalue problem
\begin{eqnarray}
H^{(0)} \Psi^{(0)} &=& 
(H^{(0)}_{\rm B}\Psi^{(0)}_{\rm B})\otimes\Psi^{(0)}_{\rm F} + 
\Psi^{(0)}_{\rm B}\otimes(H_{\rm F}^{(0)}\Psi_{\rm F}^{(0)}) \nn \\
&=& (E_{\rm B} + E_{\rm F})\Psi^{(0)} \;=\; E\Psi^{(0)}\,.
\end{eqnarray}
Let us consider below 
the bosonic and the fermionic sectors, separately.

To begin, we would like to consider the eigenvalue problem for 
the bosonic 
zero-mode Hamiltonian (\ref{b-ham}). 
The condition $H^{(0)}_{\rm B}\Psi_{\rm B}^{(0)} 
= E_{\rm B}\Psi_{\rm B}^{(0)}$ can be easily
solved since the bosonic Hamiltonian describes a collection of 
bosonic harmonic oscillators.

Now, we shall introduce annihilation operators 
\begin{eqnarray}
a_{0}^{\s I} &=& \frac{i}{\sqrt{2}}\sqrt{\frac{3}{| \mu |}}
\left(P_0^{\s I} - i \frac{\mu}{3}X_0^{\s I}\right)\,, \qquad (I=1,\,2,\,3)\,,
\\
a_{0}^m &=& \frac{i}{\sqrt{2}}\sqrt{\frac{6}{| \mu |}}
\left(P_0^m - i \frac{\mu}{6}X_0^m\right)\,, \qquad 
(m = 4,\,\ldots,\,7)\,,
\\
b_0 &=& \frac{i}{\sqrt{2}}\sqrt{\frac{6}{|\mu|}}
\left(\bar{\mathcal{P}}_0 - i \frac{\mu}{6}Z_0\right)\,, 
\end{eqnarray}
and creation operators 
\begin{eqnarray}
a_{0}^{{\s I}\,\dagger} &=& - \frac{i}{\sqrt{2}}\sqrt{\frac{3}{|\mu|}}
\left(P_0^{\s I} + i \frac{\mu}{3}X_0^{\s I}\right)\,, \qquad 
(I=1,\,2,\,3)\,,  \\
a_{0}^{m\, \dagger} &=& - \frac{i}{\sqrt{2}}\sqrt{\frac{6}{|\mu|}}
\left(P_0^m + i \frac{\mu}{6}X_0^m\right)\,, \qquad 
(m=4,\,\ldots,\,7)\,,\\
b_0^{\dagger} &=& - \frac{i}{\sqrt{2}}\sqrt{\frac{6}{|\mu|}}
\left(\mathcal{P}_0 + i \frac{\mu}{6}\bar{Z}_0\right)\,.
\end{eqnarray}
These operators satisfy the following relations, 
\begin{eqnarray}
[a_{0}^{\s I},\,a^{{\s J}\,\dagger}_{0}] \;=\; \del^{\s IJ}\,, \quad 
[a_{0}^{m},\,a^{n\,\dagger}_{0}] \;=\; \del^{mn}\,, \quad 
\left[ b_0,\,b^{\dagger}_0 \right] &=& 1\,.
\end{eqnarray}
By the use of the annihilation and creation operators, 
the bosonic zero-mode Hamiltonian $H^{(0)}_{\rm B}$ 
can be rewritten as 
\begin{eqnarray}
H_{\rm B}^{(0)} &=& \frac{|\mu|}{3}
\sum_{{\s I}=1}^3a^{{\s I}\,\dagger}_{0}a_{0}^{\s I} 
+ \frac{|\mu|}{6}
\left(
\sum_{m=4}^7 a^{m\,\dagger}_{0} a_{0}^m + 2 b^{\dagger}_{0} b_{0}
\right) + |\mu| \,,
\end{eqnarray}
where we note that the bosonic energy does not become zero due to the 
presence of the zero-point energy. 
This zero-point energy plays an important role 
to ensure the positive definiteness of the total zero-mode 
Hamiltonian, as we will see in the study of the fermionic sector.  

As usual treatment, we can obtain all eigenstates and eigenvalues. 
The wave functions are given as the products 
of Hermite polynomials. In particular, the ground-state  wave function is 
given by 
\begin{eqnarray}
\Psi_{{\rm B}\,0}^{(0)} &=& \left(\frac{|\mu|}{3\pi}\right)^{3/4}
\left(\frac{|\mu|}{6\pi}\right)^{3/2} \nn \\
&& \qquad \times \exp\Biggl[
-\frac{1}{2}\cdot\frac{|\mu|}{3}\sum_{{\s I}=1}^3(X_{0}^{\s I})^2 
- \frac{1}{2}\cdot\frac{|\mu|}{6}\sum_{m=4}^7
(X_{0}^m)^2 - \frac{|\mu|}{6}|Z_0|^2
\Biggr]\,,
\end{eqnarray}
whose energy eigenvalue is $|\mu|$, that is, 
\[
 H_{\rm B}^{(0)}\Psi_{{\rm B}\,0}^{(0)} \;=\; 
|\mu|\, \Psi_{{\rm B}\,0}^{(0)}\,.
\]
Thus, the bosonic zero-mode Hamiltonian 
has a lower bound as follows: 
\begin{equation}
 H_{\rm B}^{(0)} \;\geq\; |\mu|\,.
\label{b-bound}
\end{equation}

Next, we shall consider the fermionic zero-mode Hamiltonian (\ref{f-ham}).  
Here, 
we shall decompose the complex spinor $\lambda$ as 
\begin{eqnarray}
\lambda &=& \left(\frac{1 + i\Gamma_{123}}{2}\right)\lambda + 
\left(\frac{1 - i\Gamma_{123}}{2}\right)\lambda \nn \\
&\equiv& \lambda^{\uparrow} + \lambda^{\downarrow}
\,.
\end{eqnarray}
By definition, $\lambda^{\uparrow,\downarrow}$ should satisfy 
\begin{eqnarray}
 i\Gamma_{123}\lambda^{\uparrow,\downarrow} \;=\; \pm \lambda^{\uparrow,\downarrow}\,.
\end{eqnarray}
Now, we can rewrite the fermionic Hamiltonian 
as 
\begin{eqnarray}
H_{\rm F}^{(0)} &=& - \frac{\mu}{4}\left(
\lambda^{\uparrow}_{0\al}
\frac{\partial}{\partial\lambda^{\uparrow}_{0\al}} 
- \lambda^{\downarrow}_{0\al}
\frac{\partial}{\partial\lambda^{\downarrow}_{0\al}}
\right)\,.
\end{eqnarray}
Let us consider the zero-energy condition $H^{(0)}_{\rm F}\Psi_{\rm
F}^{(0)}=0$. 
For example, $\Psi^{(0)}_{{\rm F},0} = {\rm const}.$ is a trivial
solution. A pair of $\lambda^{\up}$ and $\lambda^{\down}$ is 
also a solution 
\[
 \Psi^{(0)}_{{\rm F},0} \;=\; \lambda^{\uparrow}_{0\al}
\lambda^{\downarrow}_{0\beta}\,.
\]
Similarly, the following states 
\begin{eqnarray} 
\lambda^{\uparrow}_{0\al_1}\lambda^{\downarrow}_{0\beta_1}
\lambda^{\uparrow}_{0\al_2}\lambda^{\downarrow}_{0\beta_2}\,, \quad 
\lambda^{\uparrow}_{0\al_1}\lambda^{\downarrow}_{0\beta_1}
\lambda^{\uparrow}_{0\al_2}\lambda^{\downarrow}_{0\beta_2}
\lambda^{\uparrow}_{0\al_3}\lambda^{\downarrow}_{0\beta_3}\,, \quad 
\lambda^{\uparrow}_{0\al_1}\lambda^{\downarrow}_{0\beta_1}
\lambda^{\uparrow}_{0\al_2}\lambda^{\downarrow}_{0\beta_2}
\lambda^{\uparrow}_{0\al_3}\lambda^{\downarrow}_{0\beta_3}
\lambda^{\uparrow}_{0\al_4}\lambda^{\downarrow}_{0\beta_4}\,,
\label{fe0}
\end{eqnarray}
also represent zero-energy states. When we note that 4 
components of $\lambda_{\al}^{\uparrow}$ or $\lambda_{\al}^{\downarrow}$ 
are non-zero, the total number of the zero-energy states is 
$1 + 16 + 36 + 16 + 1 = 70$.    
As a result, we can represent the zero-energy ground-state 
wave function of the
fermionic zero-mode Hamiltonian as the linear combination of these states
\begin{eqnarray}
\Psi^{(0)}_{\rm F,0} &=& 
c_0 + c_{2}\lambda^{\uparrow}_{0\al_1}\lambda^{\downarrow}_{0\beta_1} 
+ c_{4} \lambda^{\uparrow}_{0\al_1}\lambda^{\downarrow}_{0\beta_1}
\lambda^{\uparrow}_{0\al_2}\lambda^{\downarrow}_{0\beta_2}
+ c_{6} \lambda^{\uparrow}_{0\al_1}\lambda^{\downarrow}_{0\beta_1}
\lambda^{\uparrow}_{0\al_2}\lambda^{\downarrow}_{0\beta_2}
\lambda^{\uparrow}_{0\al_3}\lambda^{\downarrow}_{0\beta_3} \nn \\
&& + c_{8} \lambda^{\uparrow}_{0\al_1}\lambda^{\downarrow}_{0\beta_1}
\lambda^{\uparrow}_{0\al_2}\lambda^{\downarrow}_{0\beta_2}
\lambda^{\uparrow}_{0\al_3}\lambda^{\downarrow}_{0\beta_3}
\lambda^{\uparrow}_{0\al_4}\lambda^{\downarrow}_{0\beta_4}\,.
\label{state0}
\end{eqnarray} 
Moreover, we can construct nonzero-energy states containing even
number of spinors: 
\begin{itemize}
 \item a state with the energy $-\mu$, 
\begin{eqnarray}
 \lambda_{0\al}^{\uparrow}\lambda_{0\beta}^{\uparrow}
\lambda_{0\gamma}^{\uparrow}
\lambda_{0\delta}^{\uparrow}\,, 
\label{fe1}
\end{eqnarray}
 \item $6+16+6 = 28$ states with the energy $-\mu/2$,
\begin{eqnarray}
 \lambda^{\uparrow}_{0\al}\lambda^{\uparrow}_{0\beta}\,, \quad 
\lambda^{\uparrow}_{0\al}\lambda^{\uparrow}_{0\beta}
\lambda_{0\al_1}^{\uparrow}\lambda_{0\beta_1}^{\downarrow}
\,, \quad 
\lambda^{\uparrow}_{0\al}\lambda^{\uparrow}_{0\beta}
\lambda_{0\al_1}^{\uparrow}\lambda_{0\beta_1}^{\downarrow}
\lambda_{0\al_2}^{\uparrow}\lambda_{0\beta_2}^{\downarrow}\,,
\end{eqnarray}
\item $6+16+6 = 28$ states with the energy $+ \mu/2$,
\begin{eqnarray}
 \lambda^{\downarrow}_{0\al}\lambda^{\downarrow}_{0\beta}\,, \quad 
\lambda^{\downarrow}_{0\al}\lambda^{\downarrow}_{0\beta}
\lambda_{0\al_1}^{\uparrow}\lambda_{0\beta_1}^{\downarrow}
\,, \quad 
\lambda^{\downarrow}_{0\al}\lambda^{\downarrow}_{0\beta}
\lambda_{0\al_1}^{\uparrow}\lambda_{0\beta_1}^{\downarrow}
\lambda_{0\al_2}^{\uparrow}\lambda_{0\beta_2}^{\downarrow}\,,
\end{eqnarray}
\item a state with the energy $+ \mu$,
 \begin{eqnarray}
 \lambda_{0\al}^{\downarrow}\lambda_{0\beta}^{\downarrow}
\lambda_{0\gamma}^{\downarrow}
\lambda_{0\delta}^{\downarrow}\,.
\label{state1}
\end{eqnarray}
\end{itemize}

We can also construct the fermionic sector of the supergravity 
multiplet as follows: 
\begin{itemize}
 \item $4+4 =8$ fermionic states with the energy $- 3\mu/4$,
\begin{eqnarray}
\lambda_{0\al}^{\uparrow}\lambda_{0\beta}^{\uparrow}
\lambda_{0\gamma}^{\uparrow}
\,, \quad  
\lambda_{0\al}^{\uparrow}\lambda_{0\beta}^{\uparrow}
\lambda_{0\gamma}^{\uparrow}
\lambda_{0\al_1}^{\uparrow}\lambda_{0\beta_2}^{\downarrow}\,,
\label{f-state1}
\end{eqnarray}
 \item $4 + 24 + 24 + 4 = 56 $ fermionic states with the energy $- \mu/4$, 
\begin{eqnarray}
\lambda_{0\al}^{\uparrow}\,, \quad \lambda_{0\al}^{\uparrow}
\lambda_{0\al_1}^{\uparrow}\lambda_{0\beta_1}^{\downarrow}\,, \quad 
\lambda_{0\al}^{\uparrow}
\lambda_{0\al_1}^{\uparrow}\lambda_{0\beta_1}^{\downarrow}
\lambda_{0\al_2}^{\uparrow}\lambda_{0\beta_2}^{\downarrow}
\,, \quad 
\lambda_{0\al}^{\uparrow}
\lambda_{0\al_1}^{\uparrow}\lambda_{0\beta_1}^{\downarrow}
\lambda_{0\al_2}^{\uparrow}\lambda_{0\beta_2}^{\downarrow}
\lambda_{0\al_3}^{\uparrow}\lambda_{0\beta_3}^{\downarrow}\,,
\end{eqnarray}
\item $4 + 24 + 24 + 4 = 56 $ fermionic states with the energy $+ \mu/4$, 
\begin{eqnarray}
\lambda_{0\al}^{\downarrow}\,, \quad 
\lambda_{0\al}^{\downarrow}\lambda_{0\al_1}^{\uparrow}
\lambda_{0\beta_1}^{\downarrow}\,, \quad 
\lambda_{0\al}^{\downarrow}
\lambda_{0\al_1}^{\uparrow}\lambda_{0\beta_1}^{\downarrow}
\lambda_{0\al_2}^{\uparrow}\lambda_{0\beta_2}^{\downarrow}
\,, \quad 
\lambda_{0\al}^{\downarrow}
\lambda_{0\al_1}^{\uparrow}\lambda_{0\beta_1}^{\downarrow}
\lambda_{0\al_2}^{\uparrow}\lambda_{0\beta_2}^{\downarrow}
\lambda_{0\al_3}^{\uparrow}\lambda_{0\beta_3}^{\downarrow}\,,
\end{eqnarray}
 \item $4+4 =8$ fermionic states with the energy $+ 3\mu/4$,
\begin{eqnarray}
\lambda_{0\al}^{\downarrow}\lambda_{0\beta}^{\downarrow}
\lambda_{0\gamma}^{\downarrow}
\,, \quad  
\lambda_{0\al}^{\downarrow}\lambda_{0\beta}^{\downarrow}
\lambda_{0\gamma}^{\downarrow}
\lambda_{0\al_1}^{\uparrow}\lambda_{0\beta_2}^{\downarrow}\,.
\label{f-state4}
\end{eqnarray}
\end{itemize}
The resulting spectrum is listed 
in Table.\,\ref{spec:tab}. In the pp-wave case, these states are 
splitting with an energy difference $\mu/4$. 
The resulting $2(1+28+35)=128$ bosonic and the 
$2(8+56)=128$ fermionic states 
correspond to the bosonic and the fermionic sectors of the 
supergravity multiplet in the flat case.
It is confirmed that the above bosonic and the fermionic states
correspond to the supergravity multiplet in the flat case 
by noting that such states appear in the expansion of the ground-state  
wave function in the flat case as 
$\lambda_{\al}\lambda_{\beta}\ldots = (\lambda^{\uparrow}_{\al} +
\lambda_{\al}^{\downarrow})
(\lambda_{\beta}^{\uparrow} + \lambda_{\beta}^{\downarrow})\ldots$. 
These are the coefficients in front of the plane-wave in the flat case. 
If we take the $\mu \rightarrow 0$ limit, then 
the above states with non-zero energies should have zero energies, that is, 
all of them should be degenerate. 
As a result, a $128 + 128$ supergravity multiplet 
at the rest frame 
is recovered. 

Also, the resulting spectrum has the symmetry under 
$\mu \rightarrow - \mu$ as noted in Ref.\,\cite{Malda}. 
Moreover, the spectrum resembles that of superstring theories.  
That is, $1 + 28 + 35$ or $8 + 56$ looks like the 
spectrum of NS-NS, R-R or R-NS sector, although we cannot give any 
definite physical interpretation. 

Furthermore, the fermionic Hamiltonian is not positive definite, 
but it is bounded as 
\begin{equation} 
- \mu \;\leq\; H_{\rm F}^{(0)} \;\leq\; + \mu\,.  
\end{equation}
The bosonic zero-mode Hamiltonian has a lower bound as denoted 
in Eq.\,(\ref{b-bound}) due to the presence of 
the zero-point energy $|\mu|$.  
Thus, we can easily find that  the total zero-mode 
Hamiltonian is positive definite 
\[
 H^{(0)} \;\geq\; 0.
\]  
\begin{table}
\begin{center}
 \begin{tabular}{c||c|c|c|c|c|c|c|c|c}
\hline 
Energy & $-\mu$ & $-\frac{3}{4}\mu$ & $-\frac{1}{2}\mu$ &  $-\frac{1}{4}\mu$
& 0 & $+\frac{1}{4}\mu$ & $+\frac{1}{2}\mu$ & $+\frac{3}{4}\mu$ 
& $+\mu$  \\ 
\hline 
Boson & 1 & & 28 & & 35 + 35 &  & 28 &  & 1 \\
Fermion & & 8 &  & 56 & & 56 &  & 8 &  \\  
\hline
 \end{tabular}
\caption{Spectrum of the fermionic zero-mode Hamiltonian:
\footnotesize The numbers of the states at each energy level are
 listed. 
The above states form a supermultiplet under the action
 of the kinematical supersymmetries.}
\label{spec:tab}
\end{center}
\end{table}
Thus, the zero-energy ground state of the total zero-mode Hamiltonian can be 
uniquely constructed 
as\footnote{We have fixed $\mu$ positive. If $\mu$ is negative, then 
$\lambda^{\uparrow}$'s are replaced by $\lambda^{\downarrow}$'s. } 
\begin{eqnarray}
\Psi^{(0)}_{0} &=& \Psi_{{\rm B},\mu}^{(0)} \otimes \Psi^{(0)}_{{\rm F},-\mu} \nn \\
&=& \left(\frac{\mu}{3\pi}\right)^{3/4}
\left(\frac{\mu}{6\pi}\right)^{3/2}
 \lambda_{0\al}^{\uparrow}\lambda_{0\beta}^{\uparrow}
\lambda_{0\gamma}^{\uparrow}
\lambda_{0\delta}^{\uparrow}\, \nn \\
&& \qquad \times\exp\left(
-\frac{1}{2}\cdot\frac{\mu}{3}\sum_{{\s I}=1}^3(X_{0}^{\s I})^2 
- \frac{1}{2}\cdot\frac{\mu}{6}
\sum_{m=4}^7(X_{0}^m)^2 - \frac{\mu}{6}|Z_0|^2\right)\,,
\label{ground}
\end{eqnarray}
which satisfies 
\[
 H^{(0)}\Psi^{(0)}_{0} \;=\; 0\,.
\]
Also, we can directly see that 
the above ground-state  wave function 
is supersymmetric, i.e., $Q_{\al}^{(0)}\Psi_0^{(0)} 
= Q_{\al}^{(0)\dagger}\Psi_0^{(0)} = 0$ when the zero-modes 
of the dynamical supercharges are rewritten 
in terms of annihilation and creation operators as follows: 
\begin{eqnarray}
 Q_{\al}^{(0)} &=& \left(\e^{\frac{\mu}{12}\Gamma_{123}\tau}\right)_{\al\beta}
\Biggl[
i\sqrt{2}\left( \sqrt{\frac{\mu}{3}}a_0^{{\s I}\dagger}
(\Gamma_{\s I})_{\beta\gamma} + \sqrt{\frac{\mu}{6}}a_0^{m\dagger}(\Gamma_m)_{\beta\gamma}
\right)\lambda_{0\gamma}^{\uparrow}  \\ 
&& -i\sqrt{2}\left(
\sqrt{\frac{\mu}{3}}a_0^{\s I}(\Gamma_{\s I})_{\beta\gamma} 
+ \sqrt{\frac{\mu}{6}}a_0^m(\Gamma_m)_{\beta\gamma}
\right)\lambda_{0\gamma}^{\downarrow} - 2 \sqrt{\frac{\mu}{6}}\left(
b_0\frac{\partial}{\partial\lambda_{0\gamma}^{\uparrow}} 
- b_0^{\dagger}\frac{\partial}{\partial\lambda_{0\gamma}^{\downarrow}}
\right)
\Biggr]\,, \nn \\ 
Q_{\al}^{(0)\dagger} &=& \left(\e^{-\frac{\mu}{12}\Gamma_{123}\tau}\right)_{\al\beta}
\Biggl[
i\sqrt{2}\left( \sqrt{\frac{\mu}{3}}a_0^{{\s I}}
(\Gamma_{\s I})_{\beta\gamma} + \sqrt{\frac{\mu}{6}}a_0^{m}(\Gamma_m)_{\beta\gamma}
\right)\lambda_{0\gamma}^{\downarrow}  \\ 
&& -i\sqrt{2}\left(
\sqrt{\frac{\mu}{3}}a_0^{{\s I}\dagger}(\Gamma_{\s I})_{\beta\gamma} 
+ \sqrt{\frac{\mu}{6}}a_0^{m\dagger}(\Gamma_m)_{\beta\gamma}
\right)\lambda_{0\gamma}^{\uparrow} - 2 \sqrt{\frac{\mu}{6}}\left(
b_0^{\dagger}\frac{\partial}{\partial\lambda_{0\gamma}^{\downarrow}} 
- b_0\frac{\partial}{\partial\lambda_{0\gamma}^{\uparrow}}
\right)
\Biggr]\,. \nn 
\end{eqnarray}
One can see that the ground-state wave function (\ref{ground}) 
is a singlet under the $SO(3)\times SO(6)$ as shown in Appendix \ref{ang}.   

Now, we shall recall that the algebra of 
kinematical supersymmetries can be represented by\footnote{The 
factor $i$ can be absorbed into the infinitesimal 
parameter of the kinematical supersymmetry transformation.} 
\begin{eqnarray}
Q^- \;=\; \lambda_0\,, \quad 
Q^{-\dagger} \;=\; \frac{\partial}{\partial\lambda_0}\,.
\end{eqnarray}
We can see that 
the above bosonic and the fermionic states form a supermultiplet 
under the action of the kinematical supersymmetries. 
When we decompose the representation of the 
kinematical supercharges as 
\begin{eqnarray}
 Q^{-} &=& Q^{-\,\uparrow} + Q^{-\,\downarrow} \,, \nn \\
 Q^{-\dagger} &=& (Q^{-\,\uparrow})^{\dagger} 
+ (Q^{-\,\downarrow})^{\dagger}\,,\nn 
\end{eqnarray}
$Q^{-\,\uparrow}$ and $(Q^{-\,\downarrow})^{\dagger}$ 
($Q^{-\,\downarrow}$ and $(Q^{-\,\uparrow})^{\dagger}$ )
lower (raise) the fermionic energy. 
It should be remarked that the states with different energies  
belong to a multiplet because 
the supercharges do not commute with the Hamiltonian. 

The story does not end here and we have to discuss the contribution of 
nonzero-modes (i.e., off-diagonal elements of the matrix variables), 
which will be studied in the next section. 

\section{Nonzero-mode Ground-state Wave Function}

In this section, we discuss the nonzero-mode Hamiltonian, 
its ground-state wave function 
and evaluate the $L^2$-norm of the ground state. 

To begin, we shall consider the abelian part of 
the nonzero-mode Hamiltonian. According to the similar discussion to that
on the zero modes, 
we can easily show the following bounds: 
\begin{eqnarray}
 H_{\rm B}^{\rm C} \;\geq\; |\mu|(N-1) \,, \quad 
- |\mu|(N -1) \;\leq\; H_{\rm F}^{\rm C} \; \leq \; + |\mu|(N -1)\,,
\end{eqnarray}
where the lowest state of $H_{\rm F}^{\rm C}$ is represented by 
\[
 \prod_{A=1}^{N-1}\lambda^{\up}_{A\al}\lambda^{\up}_{A\beta}
\lambda^{\up}_{A\gamma}\lambda^{\up}_{A\delta}\,.
\]
Recalling that matrix variables are restricted 
to the Cartan subalgebra in the IR region, 
we can say that the nonzero-mode Hamiltonian is positive definite 
in the same way as the zero-mode Hamiltonian.  
Also, the associated ground-state wave function in the IR region
is normalizable.

We would like to consider the ground-state wave function 
of the nonzero-mode Hamiltonian including the interactions of 
non-diagonal elements of matrix variables.  
For this purpose, 
we have to solve two differential 
equations $Q\Psi = Q^{\dagger}\Psi = 0$, but it is quite difficult to 
solve them fully. In particular, the bosonic degrees of freedom for the 
$SO(6)$ part seem difficult to study.
So let us truncate the bosonic 
variables $X^i$ and $Z$ into $X^{\s I}~(I=1,2,3)$ ($SO(3)$
part) 
and also the 4-component spinors $\lambda^{\up}$ and $\lambda^{\down}$ 
into the 2-component ones, respectively. 
The truncation of the fermionic degrees is needed 
for the cancellation of the zero-point energy in the abelian parts 
of the nonzero-mode Hamiltonian. Such truncations modify the bounds of
the abelian parts in bosonic and in fermionic nonzero-mode Hamiltonians 
respectively as 
\begin{eqnarray}
 H^{\rm C}_{\rm B} &\geq& \frac{1}{2}|\mu|(N -1)\,, \quad 
- \frac{1}{2}|\mu|(N -1) \;\leq\; H^{\rm C}_{\rm F} \;\leq\; 
+\frac{1}{2}|\mu|(N -1)\,, 
\end{eqnarray}
and the lowest state of the fermionic zero-mode Hamiltonian 
becomes 
\[
\prod_{n=1}^{N-1}\lambda^{\up}_{A\al}\lambda^{\down}_{A\beta}. 
\]
Therefore the above truncation preserves the positive definiteness 
in the IR region.  

Now, the associated differential equations become simple forms  
\begin{eqnarray}
Q_{\al}\Psi &=& \left(\e^{\frac{\mu}{12}\Gamma_{123}\tau}\right)_{\al\beta}
\Biggl[
\left(
\frac{\partial}{\partial X^{\s I}_{A}} - \frac{\mu}{3}X^{\s I}_A 
- \frac{1}{2}f_{ABC}\epsilon_{IJK}X_{B}^{J}X_C^{K} 
\right)\cdot(-i(\Gamma_{\s I})_{\beta\gamma}\lambda^{\uparrow}_{A\gamma}) 
\nn \\
&& \quad + 
\left(
\frac{\partial}{\partial X^{\s I}_A} + \frac{\mu}{3}X_A^{\s I} 
+ \frac{1}{2}f_{ABC}\epsilon_{IJK}X^{\s J}_{B}X^{\s K}_{C}
\right)\cdot
(-i(\Gamma_{\s I})_{\beta\gamma}\lambda^{\downarrow}_{A\gamma})
\Biggr]\Psi  \label{diff1}
\\
&=& 0\,.\nn
\end{eqnarray} 
and 
\begin{eqnarray}
Q_{\al}^{\dagger}\Psi &=& 
-\left(\e^{-\frac{\mu}{12}\Gamma_{123}\tau}\right)_{\al\beta}
\Biggl[
\left(
\frac{\partial}{\partial X^{\s I}_{A}} + \frac{\mu}{3}X^{\s I}_A 
+ \frac{1}{2}f_{ABC}\epsilon_{IJK}X_{B}^{J}X_C^{K} 
\right)\cdot\left(-i(\Gamma_{\s I})_{\beta\gamma}
\frac{\partial}{\partial\lambda^{\uparrow}_{A\gamma}}
\right) 
\nn \\
&& \quad + 
\left(
\frac{\partial}{\partial X^{\s I}_A} - \frac{\mu}{3}X_A^{\s I} 
- \frac{1}{2}f_{ABC}\epsilon_{IJK}X^{\s J}_{B}X^{\s K}_{C}
\right)\cdot
\left(-i(\Gamma_{\s I})_{\beta\gamma}\frac{\partial}{\partial
\lambda^{\downarrow}_{A\gamma}}\right)
\Biggr]\Psi  \label{diff2}
\\
&=& 0\,.\nn
\end{eqnarray} 
We can solve the differential equations (\ref{diff1}) and (\ref{diff2}), 
and obtain a special solution for the wave function 
of nonzero-modes
\begin{eqnarray}
 \Psi_{\rm G} &=& 
\prod_{A=1}^{N^2-1}
\lambda^{\uparrow}_{A\al}\lambda^{\uparrow}_{A\beta}
\Psi_{\rm B}\,, \\
\Psi_{\rm B} &=&  \left(\frac{\mu}{3\pi}\right)^{3 ( N^2 -1 )/4}   
\exp
\Biggl[
- \frac{1}{2}\cdot\frac{\mu}{3}\sum_{{I}=1}^3\sum^{N^2-1}_{A=1}(X_A^{I})^2 
\nn \\
&& \qquad + \frac{1}{3!}\sum_{{I,J,K}=1}^3\sum^{N^2-1}_{{A,B,C}=1}
f_{ABC}\epsilon_{IJK} X_A^{I} X_B^{J} X_C^{K}  
\Biggr]\,. 
\end{eqnarray}
This solution is supersymmetric and has zero energy. 
Moreover, we can easily see that this solution is an $SO(3)\times
SO(6)$ singlet and satisfies the constraint condition
(\ref{constraint}).  
However the function $\Psi_{\rm B}$ would not be square-integrable.
In order to discuss the normalizability of this solution, 
let us set $N=2$ (i.e., $SU(2)$ matrix model) for simplicity and 
evaluate the $L^2$-norm of the above ground-state  wave function 
$||\Psi_{\rm G}||$ as 
\begin{eqnarray}
||\Psi_{\rm G}||^2 &\equiv& \int\prod_{I,A}dX^{I}_{A}
\Psi_{\rm G}^{\dagger}\Psi_{\rm G} \nn \\
&=& \int^{\infty}_{-\infty}\!\!dX^1_1
\int^{\infty}_{-\infty}\!\!dX^2_1\cdots
\int^{\infty}_{-\infty}\!\!dX^3_3 \,\,\,|\Psi_B|^2 \nn \\
&=& \left(\frac{\mu}{3\pi}\right)^{9/2}
\times I\,,
\end{eqnarray}
where the $I$ is defined by 
\begin{eqnarray}
I &=&\int^{\infty}_{-\infty}\!\!dX^1_1
\int^{\infty}_{-\infty}\!\!dX^2_1\cdots
\int^{\infty}_{-\infty}\!\!dX^3_3\,\,\,
e^{2f_0}e^{2f_1}\,, \\
f_0 &\equiv& - \frac{1}{2}\cdot\frac{\mu}{3}\sum^{3}_{A,I=1}(X_A^{I})^2 \nn\\ 
&=& - \frac{1}{2}\cdot\frac{\mu}{3}\Bigl[
(X_1^1)^2+(X_1^2)^2+(X_1^3)^2
+(X_2^1)^2 \nn \\
&& \qquad +(X_2^2)^2+(X_2^3)^2
+(X_3^1)^2+(X_3^2)^2+(X_3^3)^2
\Bigr]\,,\nonumber\\
f_1 &\equiv& \frac{1}{3!}\sum^{3}_{A,B,C, {I,J,K}=1}
\epsilon_{ABC}\epsilon_{IJK} X_A^{I} X_B^{J} X_C^{K} \nn \\
&=& -X_1^3X_2^2X_3^1+X_1^2X_2^3X_3^1+X_1^3X_2^1X_3^2 \nn \\
&& \qquad -X_1^1X_2^3X_3^2-X_1^2X_2^1X_3^3+X_1^1X_2^2X_3^3\,. \nn
\end{eqnarray}
Here, we shall expand the $f_1$ part formally and 
evaluate the integral as follows: 
\begin{eqnarray}
I=\left(\frac{3\pi}{\mu}\right)^{9/2}
\sum^{\infty}_{n=0}\frac{(2n+1)!}{n!}
\left[\frac{1}{4}\left(\frac{3}{\mu}\right)^3\right]^n\,.
\label{guess}
\end{eqnarray}
This formula is obtained by calculating several terms with lower orders 
in the expansion by using the Mathematica. 
Though we have no analytical proof, 
we guess this formula (\ref{guess}) holds for general terms. 
As a result, we can evaluate the square integral of the 
bosonic ground-state  as 
\begin{eqnarray}
 ||\Psi_{\rm G}||^2 &=& 1 + O(1/\mu^3)\,. 
\end{eqnarray}
That is, the normalizability condition can be at least 
expressed as the asymptotic series 
which can be approximated with any amount of accuracy
in the large $\mu$ limit.  
In fact, the convergence radius 
of this series is zero and in this sense we can say that 
this series is an asymptotic series. 
The $n$th partial sum of this asymptotic series does not 
converge as $n\rightarrow \infty$. However when $\mu$ is 
sufficiently large,  
we can take an $n$th partial sum for an appropriate finite $n$ 
and calculate an approximate value of $I$.
In this sense, we can write $I$ as
\begin{eqnarray}
I\sim 4\pi^{9/2}e^{-\left(\frac{\mu}{3}\right)^3}
\int^{\left(\frac{\mu}{3}\right)^{3/2}}_0\!\!dt\,\,e^{t^2}
-2\pi^{9/2}\left(\frac{\mu}{3}\right)^{-3/2}\,.\nonumber
\end{eqnarray}
Thus the corresponding wave function is not normalizable 
in the usual sense.

We note that the $n=0$ part corresponds to the
kinematical contribution 
and that $n\neq 0$ parts to the dynamical contributions arising from 
non-diagonal elements of matrix variables. 
In the IR region, the matrix variables are restricted to the 
Cartan subalgebra. Therefore 
the system can be described by the abelian part of the Hamiltonian and the 
ground state is square-integrable. 
If we consider a 
situation that the system flows from the IR region to the UV one, then 
it might be understood that the dynamical effects coming 
from non-diagonal components of matrix 
variables are turned on and 
tend to spoil down the normalizability of the 
ground-state  wave function. If we take the limit $\mu\rar \infty$, then 
we could sufficiently neglect contributions of 
off-diagonal elements (dynamical effects).   
It is also noted that the above 
series tends to diverge in the flat limit $\mu \rar 0$.

Finally, we have a comment toward the full ground-state 
wave function. In the above derivation, we have truncated 
the bosonic and fermionic variables. But, 
it would be possible 
to avoid the truncation of the fermionic degrees. 
In this case, it seems that 
the numbers of the bosons and fermions are asymmetric. 
Our interpretations are as follows: First, it might be
possible to decompose the ground-state wave function 
into the bosonic and fermionic parts as $\Psi = \Psi_{\rm B}\otimes
\Psi_{\rm F}$. It is no wonder since the 
fermionic Hamiltonian contains mass term and  
the well-known argument in the flat space would not hold on the
pp-wave.  
Therefore, if we suppose that such decomposition is possible, then it 
is not necessary that the bosonic wave function is identically zero. 
Second, 
the fermionic ground state might be completely determined by the mass
term. When we express the fermionic Hamiltonian by 
$H_{\rm F} = H_{\rm F}^{\rm Yukawa} + H_{\rm F}^{\rm mass}$, 
the vacuum expectation value (VEV) is written as 
\begin{eqnarray}
(\Psi, H_{\rm F}\Psi) &=& (\Psi, H_{\rm F}^{\rm Yukawa}\Psi) + 
(\Psi_{\rm F}, H_{\rm F}^{\rm mass}\Psi_{\rm F})\,.
\end{eqnarray} 
The VEV of the Yukawa coupling term contains the term $(\Psi_{\rm B}, 
X\Psi_{\rm B})$ and hence must vanish  
due to the requirement of an $SO(3)\times SO(6)$ symmetry for the
ground state. On the other hand, it seems quite difficult to treat 
the full bosonic part. 
In particular, the treatment of the $SO(6)$ part seems hard to
analyze. 
We also guess that the solution without the truncation of
the fermionic variables could be obtained by
setting $X^m$, $Z$ to zero in the full solution. 

\section{Conclusions and Discussions}

We have considered the ground-state  wave function of the 
supermembrane on the pp-wave by introducing the complex-spinor
notation ($Spin(7)\times U(1)$ formalism). The superalgebra 
of the complex supercharges has been also calculated. 

The zero-mode Hamiltonian and its ground state have been intensively studied. 
In particular, we have shown that the zero-mode Hamiltonian is positive 
definite and explicitly constructed the unique 
supersymmetric ground-state wave function as the product of ground
states of bosonic harmonic oscillators and fermionic variables.  
This wave function is square-integrable.  
The ground state for the zero-mode is splitting in contrast to that of
the flat case, and the zero-mode Hamiltonian has the 
string-like spectrum.

We have also investigated the nonzero-mode Hamiltonian and 
constructed an example of the supersymmetric ground-state wave function 
for the nonzero-mode with the truncation of variables in the $N=2$ case.  
This ground state seems non-normalizable, but its $L^2$-norm 
can be defined by an asymptotic series. This series diverges since 
its convergence radius 
is zero. The dynamical effects coming from non-diagonal 
elements of matrix variables are represented as 
corrections with order $1/\mu^3$. 
In the IR region, matrix variables are
restricted to the abelian part, 
whose ground-state wave function is normalizable.
We might expect that 
the interactions become stronger as the system flows from the
IR region to the UV one. 
It might be considered 
that these dynamical contributions tend 
to spoil down the normalizability of the ground state, at least in our 
example of the ground state. 
It would be possible to obtain the full solution without the
truncation. 
It would be nice to proceed the study of the nonzero-mode further. 
In particular, it is an interesting future problem to obtain the 
full nonzero-mode ground-state wave function. 
If this is possible, then one can diagonalize the nonzero-mode
Hamiltonian and obtain the spectrum. Such results shed light 
upon the spectrum of the supermembrane in the flat space. 
Also, the extension of our example to an arbitrary $N$ is an interesting 
problem. 

There exist other interesting future problems. 
The study of the Witten index for the ground state  
\cite{W-index} seems very interesting since 
the energy of bosons is different from that of the corresponding 
fermions because 
the supercharges do not commute with the Hamiltonian. 
It would be also nice to consider the supermembrane 
on less supersymmetric backgrounds as discussed in \cite{Lee}.    

\newpage

\vspace*{0.5cm}
\noindent 
{\bf\large Acknowledgement}

The work of K.S. is supported in part by the Grant-in-Aid from the 
Ministry of Education, Science, Sports and Culture of Japan 
($\sharp$ 14740115). 

\vspace*{0.5cm}

\appendix 

\vspace*{1cm}
\noindent 
{\large\bf  Appendix}
\section{Proof for the $SO(3)\times SO(6)$ Invariance of the Ground State}
\label{ang}

We shall prove that the ground-state 
wave function (\ref{ground}) of the zero-mode Hamiltonian is invariant 
under the $SO(3)\times SO(6)$ transformation.  
The orbital part of the total angular momenta trivially vanish 
due to the expressions (\ref{am1})-(\ref{am5}). 
The remaining problem is to show that the spin part vanishes. 

The spin part of (\ref{am1}) and (\ref{am2}) can be rewritten as 
\begin{eqnarray}
S^{\s IJ} &=& - \frac{i}{2}\lambda^{\up}_{0}\Gamma^{\s IJ}
\frac{\partial}{\partial\lambda_{0}^{\uparrow}} 
- \frac{i}{2}\lambda^{\down}_{0}\Gamma^{\s IJ}
\frac{\partial}{\partial\lambda_{0}^{\downarrow}} 
\,, \quad 
S^{mn} \;=\; - \frac{i}{2}\lambda^{\up}_{0}\Gamma^{mn}
\frac{\partial}{\partial\lambda_{0}^{\uparrow}}
- \frac{i}{2}\lambda^{\down}_{0}\Gamma^{mn}
\frac{\partial}{\partial\lambda_{0}^{\downarrow}} \,,
\end{eqnarray}
since the matrices $\Gamma^{\s IJ}$ and $\Gamma^{mn}$ commute with
$i\Gamma_{123}$ and the action of these matrices preserves the
chirality. 
The action of parts including $\lambda^{\down}$ trivially 
annihilate the state. The spin parts including $\lambda^{\up}$ replace the 
$\lambda_{0\al_0}$ with $\lambda_{\al}^{\up}(\Gamma^{{\s
IJ}})_{\al\al_0}$, 
($\lambda_{\al}^{\up}(\Gamma^{mn})_{\al\al_0}$). It should be noted 
that the matrices 
$\Gamma^{\s IJ}$ or $\Gamma^{mn}$ should map $\lambda^{\up}$ to 
other $\lambda^{\up}$ or zero since the 
$\Gamma^{\s IJ}$ and $\Gamma^{mn}$ are antisymmetric. 
Thus, the spin part leads 
to zero due to the overlap of $\lambda^{\up}$'s.   

Next, we shall consider the spin parts (\ref{am3}) and (\ref{am4}), which
are rewritten as 
\begin{eqnarray} 
S_0^{m+} &=& - \frac{1}{2}\lambda_{0}^{\up}\Gamma^m
\lambda_{0}^{\down}\,, \quad 
S_0^{m-} \;=\; - \frac{1}{2}\frac{\partial}{\partial
\lambda_{0}^{\up}}\Gamma^m\frac{\partial}{\partial\lambda_{0}^{\down}}\,.
\end{eqnarray}
We can easily see that these spins vanish on the ground state
(\ref{ground}). 

Finally, let us consider the spin part of (\ref{am5}), which is given 
by 
\begin{eqnarray}
S^{89}_0 &=& -\frac{1}{2}\lambda_0^{\up}
\frac{\partial}{\partial\lambda_0^{\up}} 
- \frac{1}{2}\lambda_0^{\down}\frac{\partial}{\partial\lambda_0^{\down}} 
+ 2\,. 
\end{eqnarray}
The part containing $\lambda^{\down}$ trivially annihilates the state.  
The action of the part written by $\lambda^{\up}$ leads to 
the eigenvalue $-2$, which is canceled with the normal-ordering constant
$+2$. 

The extension of the above proof to the nonzero-mode 
ground-state wave function discussed in the manuscript 
is a simple exercise.

\end{document}